\documentclass[aps,prb,showpacs,superscriptaddress,twocolumn]{revtex4}

\usepackage{amsmath,amssymb}
\usepackage{latexsym}
\usepackage{graphicx}
\usepackage[bf,tiny,center,indentafter]{titlesec}

\begin{document}

\title{Transient heat generation in a quantum dot under a step-like pulse bias}

\author {Wei Pei}
\affiliation{Institute of Physics, Chinese Academy of Sciences,
Beijing 100190, China}

\author{X. C. Xie}
\affiliation{International Center for Quantum Materials, Peking
University, Beijing 100871, China} \affiliation{Department of
Physics, Oklahoma State University, Stillwater, Oklahoma 74078}

\author{Qing-feng Sun}
\email{sunqf@iphy.ac.cn} \affiliation{Institute of Physics, Chinese
Academy of Sciences, Beijing 100190, China}

\email[]{sunqf@iphy.ac.cn}

\begin{abstract}
We study the transient heat generation in a quantum dot system
driven by a step-like or a square-shaped pulse bias. We find that a
periodically oscillating heat generation arises after adding the
sudden bias. One particularly surprising result is that there exists
a heat absorption from the zero-temperature phonon subsystem. Thus
the phonon population in non-equilibrium can be less than that of
the equilibrium electron-phonon system. In addition, we also
ascertain the optimal conditions for the operation of a quantum dot
with the minimum heat generation.
\end{abstract}

\pacs{65.80.-g, 44.90.+c, 73.23.-b}

\maketitle
\titlespacing{\section}{0pt}{10pt}{*4}

\section{INTRODUCTION}
With the technologic development in the micro-integration of
electronic devices, the heat generation problem has sharply emerged.
This increasing thermal generation greatly influences the
performance and stability of nanoscale electronic devices, and it
becomes a main bottleneck for the future advancement of electronic
industry. Furthermore, some familiar concepts of heat generation
that are applicable in macroscopic systems may not be valid in
nanoscale systems. So there is an imperative need to comprehend
fully the heating effect and to try to mitigate it.

Owing to difficulties in probing the heating process in
nanostructures, not much progress had been made until several
state-of-the-art experimental scenarios were conceived in recent
years.
\cite{Smit2004,Tsutsui2006,Huang2006,Huang2007,Tsutsui2008,Tsutsui2010,Oron-Carl2008,Ioffe2008}
There are both indirect
\cite{Smit2004,Tsutsui2006,Huang2006,Huang2007,Tsutsui2008,Tsutsui2010}
and direct \cite{Oron-Carl2008,Ioffe2008} methods, which enable the
evaluation of the effective local temperatures of nanoscale
junctions. It has been experimentally demonstrated that the local
heating may induce a substantial temperature increase in single
molecular junctions
\cite{Huang2006,Huang2007,Tsutsui2008,Tsutsui2010} because of the
inefficient heat dissipation. From a microscopic point of view, the
main factor of heat generation is the electron-phonon (e-p)
interaction, which transfers the ordered energy of electric current
to the disordered vibrational energy of atoms. In some recent
measurements of differential conductance, phonon-assisted tunnelling
peaks have been observed in various nano-devices (e.g.
$\mathrm{C}_{60}$, carbon nanotube, etc.),
\cite{Park2000,LeRoy2004,LeRoy2005,Sapmaz2006,Huttel2009,Leturcq2009,Steele2009}
these clearly show the existence of e-p interaction. As for the
aspect of theory, on the one hand much efforts have been made to
investigate the effects of inelastic e-p scattering on electronic
transport through different nanostructures,
\cite{Montgomery2003,Chen2004,Chen2005,Paulsson2005,GalperinJPCM2007,Rudzinski2008,Riwar2009,Tahir2010,Fang2011}
on the other hand many theoretical works have devoted to
understanding the phenomena of heat, various models and methods were
suggested as well.
\cite{Chen2003,Lazzeri2005,Horsfield2006,Sun2007,Galperin2007,Lu2007,Liu2009,Wu2009,Zhou2011,Xie2011,Gurevich2011,Chi2012}
By using the nonequilibrium Green's function (NEGF) method, a
tractable formula of heat generation has been derived.
\cite{Sun2007} But these works mainly focused on the steady-state
cases or, at most, the sinusoidal-like ac bias. In order to analyze
the transient heat generation and its dynamics, the step-like and
the square-shaped pulse biases are considered as the optimal driving
forces. \cite{Plihal2000,Maciejko2006,Xing2007} It is not only
because they can provide the unambiguous time scales and the
transient behaviours of heat generation, but also because any
information in computers is presented in bytes that consist of a
long series of \textquotedblleft0\textquotedblright
and\textquotedblleft1\textquotedblright digits, they are realized by
continuously switching on and off the sharp step pulse.

In this paper, we study the transient behaviours of heat generation
in a quantum dot (QD) system driven by a step-like or a
square-shaped pulse bias. The QD is described by the Anderson model
and the intra-dot e-p interaction is considered. By using the NEGF
method, the transient heat generation is derived. We find a sudden
bias inducing a periodically oscillating heat generation. Moreover,
it can give rise to an energy transfer from the zero-temperature
phonon subsystem to the electronic subsystem under certain
conditions. However, after the bias goes back to zero and the system
restores the equilibrium, the total heat generation at
zero-temperature is always positive because of the third law of
thermodynamics. In addition, we also discuss how to operate the QD
with as little heat generation as possible.

The rest of the paper is organized as follows. In Sec.~\ref{section_2} we
describe the model and deduce the formula of the transient heat
generation. In Sec.~\ref{section_3} we give the numerical results and
discussion. Finally, a brief conclusion is presented in Sec.~\ref{section_4}.

\section{THEORETICAL MODEL}
\label{section_2}
To begin with, we consider a single-level QD connecting to two
electrode-leads and coupling to a local vibronic phonon mode. The
whole system is described by the typical Hamiltonian: \cite{Sun2007}
\begin{eqnarray}
H &=& {\omega_q}\hat{a}_q^{\dagger}\hat{a}_q + {\lambda_q}[\hat{a}_q
+ \hat{a}_q^{\dagger}] \hat{n}_d
+ {\epsilon_d} \hat{n}_d \nonumber\\
&+& \sum_{\alpha,k}{\epsilon_{{\alpha}k}(t)}\hat{c}_{{\alpha}k}^{\dagger}\hat{c}_{{\alpha}k}
+ \sum_{\alpha,k}{t_{\alpha}}[\hat{c}_{{\alpha}k}^{\dagger}\hat{d}
+ \mathrm{H}.\mathrm{c}.].
\label{original_hamiltonian}
\end{eqnarray}
Herein, $\hat{n}_d=\hat{d}^{\dagger}\hat{d}$, $\hat{d}$ and
$\hat{c}_{{\alpha}k}$ ($\alpha=\mathrm{L},\mathrm{R}$) are electronic annihilation
operators in the QD and lead $\alpha$, $\epsilon_d$ and
$\epsilon_{{\alpha}k}(t)$ are the corresponding single-particle
energies, and $t_{{\alpha}}$ represents the hopping matrix element
between the QD and lead $\alpha$. As for the phonon subsystem,
$\hat{a}_q$ ($\hat{a}_q^{\dagger}$) is an annihilation (creation)
operator for one specific vibronic mode with frequency $\omega_q$.
The strength of the e-p interaction is described by the coupling
constant $\lambda_q$.

Due to the pulse bias, the electronic level
$\epsilon_{{\alpha}k}(t)$ in Eq.~(\ref{original_hamiltonian}) is time-dependent:
$\epsilon_{{\alpha}k}(t) = \epsilon_{{\alpha}k}+\Delta_{\alpha}(t)$,
and $\Delta_{\alpha}(t)$ represents the external time-dependent bias
potential. For the step-like pulse case, we assume that the system
is initially in equilibrium under the zero bias; at $t = 0$ a bias
$V$ is suddenly switched on and remains for $t > 0$, which drives
the system out of equilibrium, induces the electric current and then
generates heat. Following we study the transient heat generation for
$t > 0$. Hereafter we consider that the bias is applied solely on the
left lead, so $\Delta_\mathrm{R}(t) = 0$, $\Delta_\mathrm{L}(t) = 0$ for $t < 0$, and
$\Delta_\mathrm{L}(t) = V$ otherwise.

In what follows, it is rewarding to make a canonical transformation
on the Hamiltonian $\bar{H} = UHU^{\dagger}$, with the unitary
operator $U = \exp[(\lambda_q/\omega_q)(\hat{a}_q^{\dagger}-\hat{a}_q)\hat{d}^{\dagger}\hat{d}]$:
\begin{eqnarray}
\bar{H} &=& {\omega_q}\hat{a}_q^{\dagger}\hat{a}_q
+ {\tilde{\epsilon}_d}\hat{n}_d \nonumber\\
&+& \sum_{\alpha,k}{\epsilon_{{\alpha}k}(t)}\hat{c}_{{\alpha}k}^{\dagger}\hat{c}_{{\alpha}k}
+ \sum_{\alpha,k}{t_{\alpha}}[\hat{c}_{{\alpha}k}^{\dagger}\hat{d}\hat{X}
+ \mathrm{H}.\mathrm{c}.],
\label{transformed_hamiltonian}
\end{eqnarray}
where $\tilde{\epsilon}_d = \epsilon_d - \lambda_q^2/\omega_q$ is
the renormalized electronic level of the QD, and the operator $\hat{X} =
\exp[-(\lambda_q/\omega_q)(\hat{a}_q^{\dagger}-\hat{a}_q)]$. We use
the general formula of transient heat generation presented in
Ref.~\onlinecite{Sun2007}, which remains unchange under the canonical
transformation:
\begin{eqnarray}
Q(t) &=& -2\mathrm{I}\mathrm{m} \omega_q\lambda_q^2
{\int_{-\infty}^t} dt_1
e^{-\mathrm{i}\omega_q(t_1-t)} \{ \nonumber\\
&-& N_q\langle\langle{\hat{n}_d(t)|\hat{n}_d(t_1)}\rangle\rangle^r
+ \langle\langle{\hat{n}_d(t)|\hat{n}_d(t_1)}\rangle\rangle^<\}.
\label{original_heat_generation}
\end{eqnarray}
Next we replace the operator $\hat{X}$ by its expectation value
${\langle}\hat{X}{\rangle} =
\exp[-(\lambda_q/\omega_q)^2(N_q+1/2)]$, which is reasonable under
the conditions of both $t_{\alpha}\ll\lambda_q$ and
$t_{\alpha}\gg\lambda_q$. \cite{Chen2005,Mahan1990,Hewson1980} After this
approximation, the electron and the phonon subsystems are decoupled. So
we can rigorously employ the Wick's theorem to deal with the
two-particle Green's function: $
{\langle}T_c[\hat{n}_d(t)\hat{n}_d(t_1)]{\rangle}_{\bar{H}} =
{\langle}T_c[\hat{n}_d(t)]{\rangle}_{\bar{H}} {\langle}
 T_c[\hat{n}_d(t_1)]{\rangle}_{\bar{H}} -
{\langle}T_c[\hat{d}^{\dagger}(t)\hat{d}(t_1)]{\rangle}_{\bar{H}}
{\langle} T_c[\hat{d}^{\dagger}(t_1)\hat{d}(t)]{\rangle}_{\bar{H}}$.
Then the heat generation $Q(t)$ changes
into:
\begin{eqnarray}
Q(t) &=& 2\mathrm{R}\mathrm{e}\omega_q\lambda_q^2
{\int_{-\infty}^t} dt_1
e^{-\mathrm{i}\omega_q(t_1-t)} \{ \nonumber\\
&-& N_q[\widetilde{G}^r(t,t_1)\widetilde{G}^<(t_1,t)
+ \widetilde{G}^<(t,t_1)\widetilde{G}^a(t_1,t)] \nonumber\\
&+& [\widetilde{G}^<(t,t_1)\widetilde{G}^>(t_1,t)
- \widetilde{G}^<(t,t)\widetilde{G}^<(t_1,t_1)] \},
\label{derived_heat_generation}
\end{eqnarray}
where $\widetilde{G}^{r,a,<,>}(t,t_1)$ are the standard
single-particle Green's functions of the QD derived from the decoupled
Hamiltonian. $N_q = 1/[\exp(\omega_q/k_{\mathrm{B}} T)-1)]$ is the Bose
distribution function with temperature $T$. Then by considering the
zero temperature case ($T=0$), the expression of transient heat
generation reduces to:
\begin{eqnarray}
Q(t) &=& 2\mathrm{R}\mathrm{e}\omega_q\lambda_q^2
{\int_{-\infty}^t} dt_1
e^{-\mathrm{i}\omega_q(t_1-t)} \nonumber\\
&\times& [\widetilde{G}^<(t,t_1)\widetilde{G}^>(t_1,t)
- \widetilde{G}^<(t,t)\widetilde{G}^<(t_1,t_1)],
\label{reduced_heat_hamiltonian}
\end{eqnarray}
and the Green's functions $\widetilde{G}^{<,>}$ take the forms
as: \cite{Haug1998,Jauho1994}
\begin{eqnarray}
\widetilde{G}^<(t,t_1) &=& \sum_{\alpha} \mathrm{i}
\widetilde{\Gamma}_{\alpha} {\int}\frac{d\epsilon}{2\pi}
f(\epsilon)\widetilde{A}_{\alpha}(\epsilon,t)\widetilde{A}_{\alpha}^*(\epsilon,t_1) \nonumber\\
& & \times \exp[-\mathrm{i}\epsilon(t-t_1)-\mathrm{i}\int_{t_1}^tdt'\Delta_{\alpha}(t')],
\label{less_green_function}\\
\widetilde{G}^>(t,t_1) &=& \widetilde{G}^<(t,t_1)
+\widetilde{G}^r(t,t_1) - [\widetilde{G}^{r}(t,t_1)]^*.
\label{greater_green_function}
\end{eqnarray}
In Eq.~(\ref{less_green_function}), the linewidth function $\widetilde{\Gamma}_{\alpha}
= \Gamma_{\alpha} {\langle}\hat{X}{\rangle}^2$, $\Gamma_{\alpha} =
2\pi\rho_{\alpha} |t_{\alpha}|^2$, with the density of state
$\rho_{\alpha}$ of lead $\alpha$. $f(\epsilon) =
1/\{\exp[(\epsilon-\mu)/k_{\mathrm{B}}T]+1\}$ is the Fermi distribution function.
To avoid trivial details, we have introduced the quantity
$\widetilde{A}_{\alpha}(\epsilon,t)$ defined as: \cite{Haug1998,Jauho1994}
\begin{eqnarray}
\widetilde{A}_{\alpha}(\epsilon,t) &=& {\int_{-\infty}^t} dt'
\widetilde{G}^r(t,t') \nonumber\\
& & \times \exp[\mathrm{i}\epsilon(t-t') + \mathrm{i}\int_{t'}^tdt''\Delta_{\alpha}(t'')].
\label{quantity_a}
\end{eqnarray}
The retarded Green's function
$\widetilde{G}^r(t,t') = -\mathrm{i}\theta(t-t')
\exp[-\mathrm{i}\tilde{\epsilon}_d(t-t')-\widetilde{\Gamma}(t-t')]$
can be easily derived, in which
$\widetilde{\Gamma} = (\widetilde{\Gamma}_\mathrm{L} + \widetilde{\Gamma}_\mathrm{R})/2$.
It is independent of the bias potential $\Delta_{\alpha}(t)$.
Combining the Eqs.~(\ref{reduced_heat_hamiltonian}--\ref{quantity_a}),
the transient heat generation $Q(t)$ can be calculated for arbitrary
time-dependent bias $\Delta_{\alpha}(t)$. Once $Q(t)$ is known, the
total heat generation $Q_T(t)$ from $t=0$ to the time $t$ is
straightforwardly obtained as: $Q_T(t) = \int_{-\infty}^t dt' Q(t')
= \int_{0}^t dt' Q(t')$. Concretely, by taking into account of the
step-like pulse bias, $\widetilde{A}_{\alpha}(\epsilon,t)$ is to be simplified:
\begin{eqnarray}
\widetilde{A}_\mathrm{R}(\epsilon,t)
&=& \widetilde{G}^r(\epsilon)
= 1/(\epsilon-\tilde{\epsilon}_d + \mathrm{i}\widetilde{\Gamma}),
\label{righr_a}\\
\widetilde{A}_\mathrm{L}(\epsilon,t<0) &=& \widetilde{G}^r(\epsilon),
\label{left_a_time_less_than_zero}\\
\widetilde{A}_\mathrm{L}(\epsilon,t>0)
&=& \widetilde{G}^r(\epsilon+V)+[\widetilde{G}^r(\epsilon)-\widetilde{G}^r(\epsilon+V)] \nonumber\\
& & \times \exp[{\mathrm{i}(\epsilon+V-\tilde{\epsilon}_d + \mathrm{i}\widetilde{\Gamma})t}].
\label{left_a_time_greater_than_zero}
\end{eqnarray}
In the following calculation, we will set the chemical potential $\mu=0$
of the equilibrium system at $t<0$ as the zero point of energy, and
also assume the symmetric barriers $\widetilde{\Gamma}_\mathrm{L} =
\widetilde{\Gamma}_\mathrm{R}$.

\section{RESULTS AND DISCUSSION}
\label{section_3}
\begin{figure}
\includegraphics[bb = 29 24 794 545, width = 0.5\textwidth]{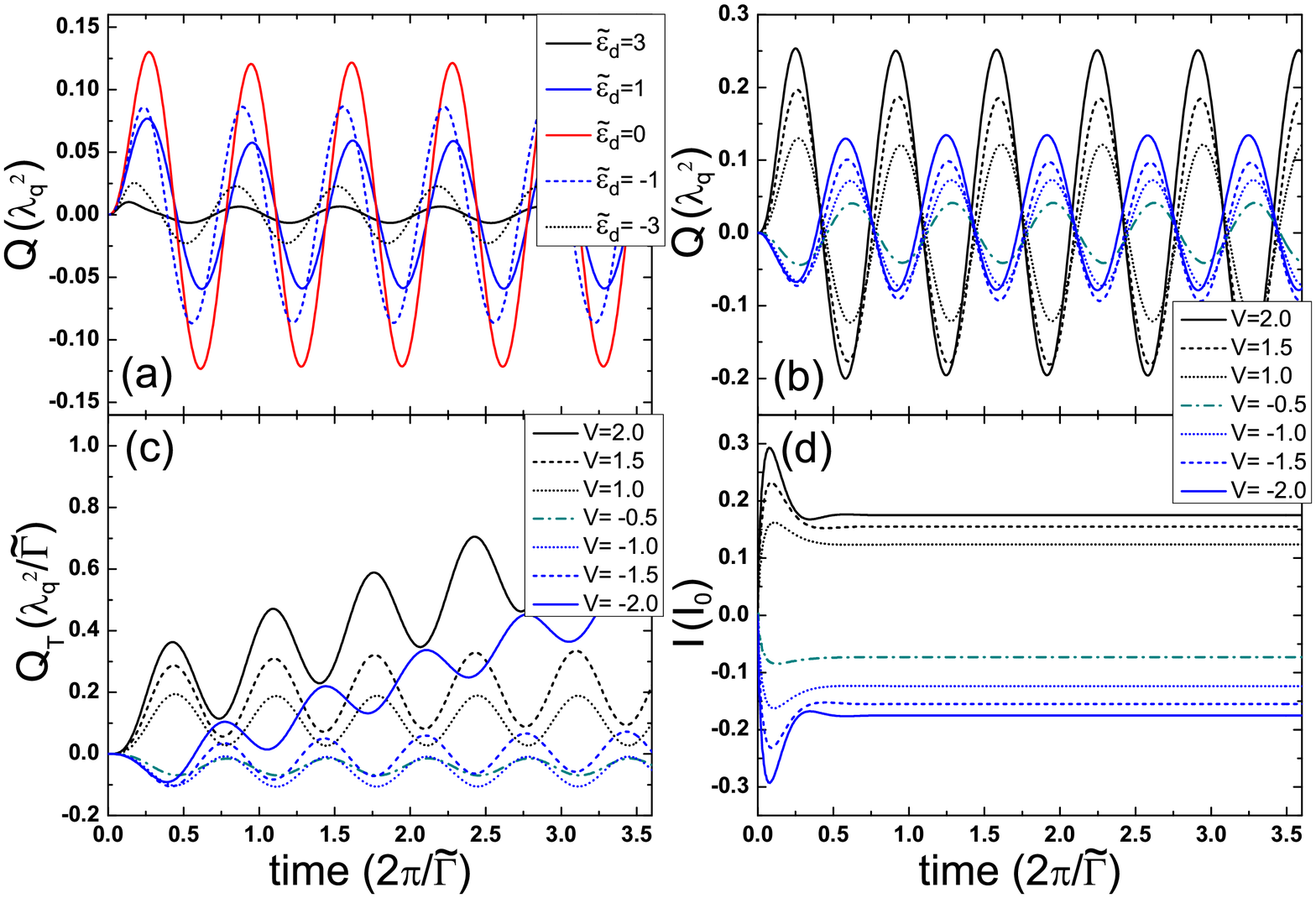}\\
\caption{(Colour online) The transient behaviours in response to a
sudden step-like pulse bias with phonon subsystem having a single
vibronic mode. (a) $Q(t)$ vs. $t$ with $V = 1$. (b), (c) and (d) are
respectively $Q(t)$, $Q_T(t)$, and $I(t)$ vs. $t$ with
$\tilde{\epsilon}_d = 0$. Other parameters are $\omega_q = 1.5$ and
$\widetilde{\Gamma} = 1$. The unit $I_0 =
e\widetilde{\Gamma}/\hbar$.}
\label{figure_1}
\end{figure}

Fig.~\ref{figure_1}(a) shows the transient heat generation $Q(t)$ versus the time
$t$ with fixed $V = 1$ and different renormalized level
$\tilde{\epsilon}_d$. While $t<0$, the bias is zero and the system
is in equilibrium, so one always has $Q(t)=0$. At $t=0$, the bias is
suddenly added, then the current $I(t)$ and the heat generation $Q(t)$
emerge at $t>0$. $Q(t)$ displays a periodically oscillating behaviour
with the period $2\pi/\omega_q$. The oscillation is undamped and
keeps all along due to only a discrete phonon model being considered
here. $Q(t)$ is positive in one half period and negative in other half
period, which means that the transporting electron periodically
emits and absorbs phonon. Besides, the oscillating amplitude is
sensitive to $\tilde{\epsilon}_d$, it is the largest while
$\tilde{\epsilon}_d=0$ because of the occurrence of resonant
tunnelling with the largest current for that $\tilde{\epsilon}_d$.
When $\tilde{\epsilon}_d$ deviates from zero, the oscillating
amplitude reduces and gradually approaches zero.

Next we fix the level $\tilde{\epsilon}_d =0$ and investigate the
behaviours of $Q(t)$ with different bias $V$. For the positive bias $V$,
$Q(t)$ is positive in the pre-half period and negative in the post-half
period. But for the negative $V$, $Q(t)$ can be negative in the
pre-half period [see the region of $0<t<\pi/\omega_q \approx
0.33(2\pi/\widetilde{\Gamma})$ in Fig.~\ref{figure_1}(b)]. This result is
surprising and indicates that the suddenly-added bias causes the
phonon to be transferred from the phonon subsystem to the electronic
subsystem at once. Notice that the temperature of the system has
been set equal to zero, so it means the phonon population can be {\sl
less} than the original zero-T value. With $|V|$ increasing, the
oscillating amplitude of $Q(t)$ also increases. For the positive-$Q$
half period, the amplitude increase is very prominent. But for the
negative-$Q$ half period, the amplitude increase is small, and even
decreases for the large $|V|$. So in the case of large $|V|$
($|V|>\omega_q$), more phonons are emitted than absorbed, which
leads to a net heat generation by the current.

\begin{figure}
\includegraphics[bb = 61 14 784 522, width = 0.5\textwidth]{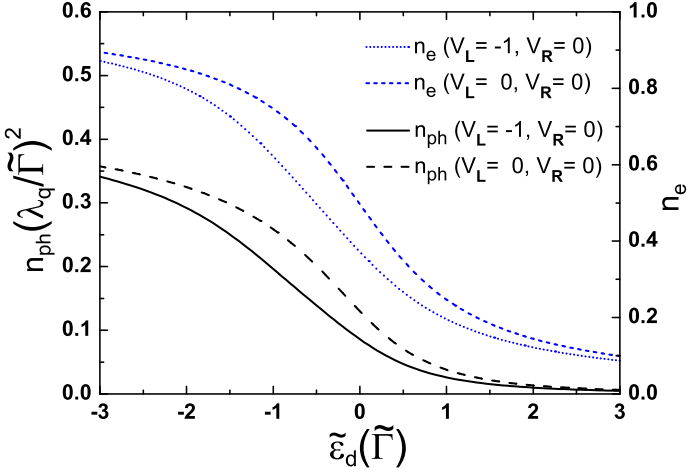}\\
\caption{(Colour online) The occupation number $n_{\mathrm{ph}}$ of phonon and
the occupation number $n_{\mathrm{e}}$ of electron on the QD at zero temperature with respect to the
renormalized electronic level $\tilde{\epsilon}_d$. Both the equilibrium-state case ($V_{\mathrm{L}}=0,V_{\mathrm{R}}=0$)
and the steady-state case ($V_{\mathrm{L}}=-1,V_{\mathrm{R}}=0$) are considered.
The parameters are set as $\omega_q=1.5$, $\widetilde{\Gamma}=1$.}
\label{figure_3}
\end{figure}

In Fig.~\ref{figure_1}(c), the total heat generation $Q_T(t)$ is shown. Because of
the oscillating $Q(t)$, $Q_T(t)$ also exhibits the oscillating
property with time $t$. For the small bias ($0<|V|<\omega_q$),
$Q_T(t)$ oscillates along a horizontal line, in which case there is
no net heat generation. But for the large bias ($|V|>\omega_q$), an
ascending trend will grow more and more noticeably, by considering the large time scale,
$Q_T(t)$ is approximately proportional to $t$. In particular, as for $-\omega_q<V<0$, the curves of $Q_T(t)$
always lie below the zero point [see the curves with $V=-0.5$ and
$-1.0$ in Fig.~\ref{figure_1}(c)], which means that at any time there is a net
phonon absorption from the zero-T phonon subsystem.
Intuitively, this seems to be impossible. But in fact, due to the
e-p interaction, the equilibrium phonon population
$n_{\mathrm{ph}}={\langle}a_q^{\dagger}(t)a_q(t){\rangle}$ at $t < 0$ is not
zero, but a finite value even if $T=0$. By using the NEGF method and
taking the same approximation as in the derivation of $Q(t)$,
$n_{\mathrm{ph}}$ can be calculated.
It is a function of $\tilde{\epsilon}_d$ behaving like the
occupation number of electron on the QD, and we will mention it
later in this paper (see Fig.~\ref{figure_3}). Then by choosing the same
parameters in Fig.~\ref{figure_1}(c), $n_{\mathrm{ph}}$ is about
$0.13\lambda_q^2/\widetilde{\Gamma}^2$. From the above results, we find
that the suddenly-applied bias can transfer a part of energy of such $n_{\mathrm{ph}}$ phonons from
the original zero-T phonon subsystem to the electronic subsystem.
Due to the system at $t>0$ being in non-equilibrium, this
transfer is not prohibited in principle. Furthermore, we have
already checked that the total heat absorption $-Q_T(t)$ from the
phonon subsystem does not exceed the value of $\omega_q
n_{\mathrm{ph}}\approx 0.20\lambda_q^2/\widetilde{\Gamma}$ in any case.

The response of transient electric current $I(t)$ to the sudden
pulse bias is also shown in Fig.~\ref{figure_1}(d),
\footnote{In the current calculation, we use the same method in Ref.~\onlinecite{Chen2005}.}
in which $I(t)$
quickly rises in a very short time. The rising time scale of $I(t)$
is about $0.1(2\pi/\widetilde{\Gamma})$. However, the rise of $Q(t)$
is within the time scale $(1/4)(2\pi/\omega_q)$. So if we set the
parameter $\widetilde{\Gamma}>0.4\omega_q$, the current $I(t)$ will
emerge much quicker than $Q(t)$. In this case, the device can finish
the operation before the occurrence of heat generation.
Additionally, in comparison with $Q(t)$, 1) $I(t)$ is without
oscillation and quickly damps to a stable value because of the
spectrum of electronic reservoirs (lead $\mathrm{L}$ and $\mathrm{R}$) being
continuous, 2) $I(t)$ has the electron-hole symmetry [e.g.
$I(t,-V,-\tilde{\epsilon}_d)= -I(t,V,\tilde{\epsilon}_d)$], but
$Q(t,-V,-\tilde{\epsilon}_d)\not= \pm Q(t,V,\tilde{\epsilon}_d)$, it
is because that the phonon subsystem is without the electron-hole symmetry.

\begin{figure}
\includegraphics[bb = 35 253 796 545, width = 0.5\textwidth]{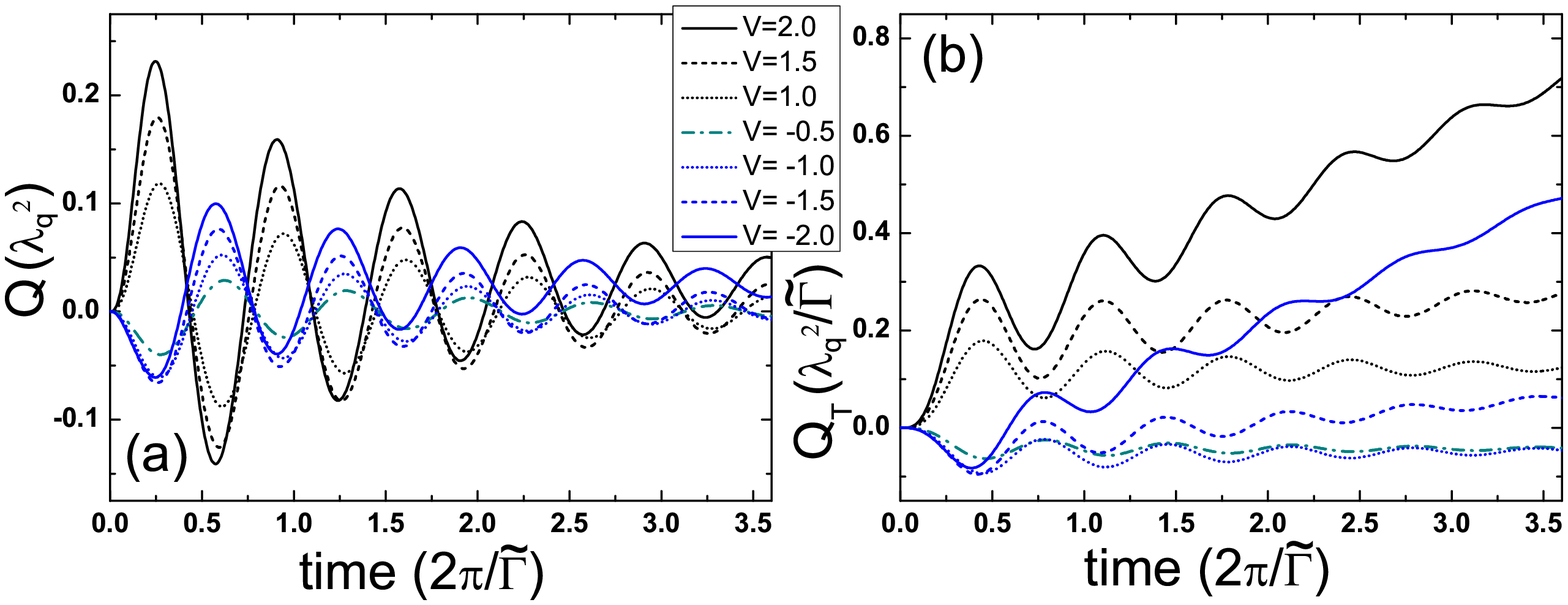}\\
\caption{(Colour online) The transient behaviours in response to a
sudden step-like pulse bias with phonon subsystem having a Lorentzian
spectrum. (a) and (b) are $Q(t)$ and $Q_T(t)$ respectively. The
parameters are $\tilde{\epsilon}_d = 0$, $\omega^0_q = 1.5$, $\sigma
= 0.1$, and $\widetilde{\Gamma} = 1$.}
\label{figure_2}
\end{figure}

Up to now, we have only considered a single isolated phonon mode
$\omega_q$ in the QD. But for a real device, the QD usually couples
to the lead and the substrate, so the phonon mode $\omega_q$ will
unavoidably acquire a small line width. This is to be
considered in the following discussion with the phonon subsystem
having a Lorentzian spectrum:
\begin{eqnarray}
A(\omega_q)
= \frac{\theta(\omega_q)}{\pi} \frac{\sigma}{(\omega_q-\omega_q^0)^2 + \sigma^2}.
\label{spectral_function}
\end{eqnarray}
Now the oscillations of $Q(t)$ and $Q_T(t)$ gradually
damp as shown in Fig.~\ref{figure_2}. $Q(t)$ damps to a finite positive value for
$|V| \geq \omega^0_q$ or almost zero for $|V|<\omega^0_q$. $Q_T(t)$
damps along a slanting line for $|V| \geq \omega^0_q$ or a
horizontal line for $|V| <\omega^0_q$. In particular, for
$-\omega^0_q <V<0$, $Q(t)\approx 0$ and $Q_T(t)$ is negative after
the system reaches a new steady state at the large time $t$ [see the
curves with $V=-0.5$ and $V=-1.0$ in Fig.~\ref{figure_2}(b)]. It means that, in the
process of applying the sudden bias, a net energy transfer from the
phonon subsystem to the electronic subsystem occurs. The phonon
population $n_{\mathrm{ph}}$ in the new steady state is less than that of the
original zero-T case and it can keep this value for a very long time
because of $Q(t)\approx 0$. To pay attention, the system is now in
the steady state, not in equilibrium. So $n_{\mathrm{ph}}$ being less than
the original zero-T value does not indicate that its temperature is
lower than $0$, and this result does not breach the third law of
thermodynamics either.

\begin{figure}
\includegraphics[bb = 219 88 609 539, width = 0.5\textwidth]{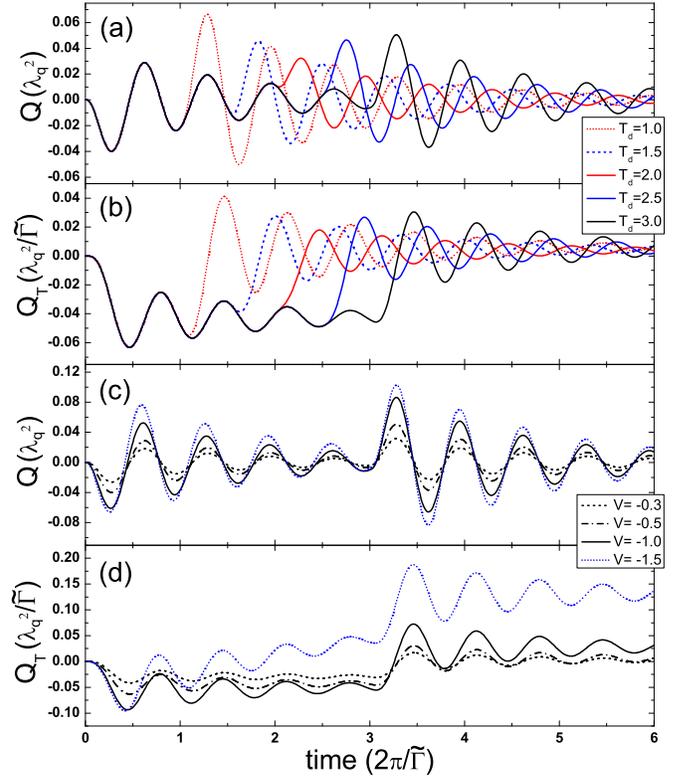}\\
\caption{(Colour online) The transient behaviours in response to a
square-shaped pulse bias with phonon subsystem having a Lorentzian
spectrum. (a) and (c) show $Q(t)$ vs. $t$, (b) and (d) are $Q_T(t)$
vs. $t$. In (a) and (b), $V=-0.5$, in (c) and (d), $T_d=3$. Other
parameters are the same as those in Fig.~\ref{figure_2}.}
\label{figure_4}
\end{figure}

In order to further explain and comprehend the phenomenon of heat
absorption from the phonon subsystem at zero temperature, we have
calculated the occupation number of zero-T phonon,
$n_{\mathrm{ph}}$, and the occupation number of electron,
$n_{\mathrm{e}}$, in both equilibrium state and steady state:
\begin{eqnarray}
n_{\mathrm{e}} &=& \langle \hat{n}_d \rangle = -\mathrm{i}\int
\widetilde{G}^<(\epsilon) d\epsilon/2\pi, \label{ne} \\
n_{\mathrm{ph}} &=& \frac{\lambda_q^2}{4\pi^2} \int \int
{d\epsilon_1} {d\epsilon_2} \nonumber\\
&\times& [\frac{\widetilde{G}^<(\epsilon_1)\widetilde{G}^>(\epsilon_2)}{(\epsilon_1-\epsilon_2-\omega_q)^2}
-
\frac{\widetilde{G}^<(\epsilon_1)\widetilde{G}^<(\epsilon_2)}{\omega_q^2}
], \label{nph}
\end{eqnarray}
where $\widetilde{G}^<(\epsilon)=\mathrm{i}\widetilde{G}^r(\epsilon)
[\sum_{\alpha}\widetilde{\Gamma}_{\alpha}f_{\alpha}(\epsilon)]
\widetilde{G}^a(\epsilon)$,
$\widetilde{G}^>(\epsilon)=-\mathrm{i}\widetilde{G}^r(\epsilon)
\{\sum_{\alpha}\widetilde{\Gamma}_{\alpha}[1-f_{\alpha}(\epsilon)]\}
\widetilde{G}^a(\epsilon)$, and
$f_{\alpha}(\epsilon)=1/\{\exp[(\epsilon-\mu_{\alpha})/k_{\mathrm{B}}T]+1\}$
($\alpha=\mathrm{L},\mathrm{R}$). See Fig.~\ref{figure_3}, the
occupation number of phonon behaves much like that of electron,
which monotonically decreases with the elevated renormalized level
of the QD, that is to say, $n_{\mathrm{ph}}$ is approximately
proportional to $n_\mathrm{e}$ while changing $\tilde{\epsilon}_d$.
At $t<0$, there is no bias across the QD
($V_{\mathrm{L}}=0,V_{\mathrm{R}}=0$), so the whole system is in
equilibrium. After applying the sudden bias, the system is driven
out of equilibrium, until a large time, enters a new steady state
($V_{\mathrm{L}}=-1,V_{\mathrm{R}}=0$). As we can see, both the
phonon population $n_{\mathrm{ph}}$ and the electron population
$n_{\mathrm{e}}$ experience an obvious falling down from the
equilibrium-state case to the steady-state case, it means that a
part of zero-T phonons have indeed run away, which results in the
heat absorption. To be specific, for $\tilde{\epsilon}_d=0$, the
change of the average phonon number at zero temperature is about
$0.044\lambda_q^2/\widetilde{\Gamma}^2$, and the corresponding
phonon energy is $0.066\lambda_q^2/\widetilde{\Gamma}$. As a matter
of fact, the net heat absorption $-Q_T$ in real case will be a
little less than this value, in that the electric current can also
transfer energy to the phonon subsystem as an offset [see the
approached level of $Q_T$ with $V=-1.0$ in Fig.~\ref{figure_2}(b)].

Finally, we deal with the square-shaped pulse bias $\Delta_\mathrm{L}(t)$, in which
case $\Delta_\mathrm{L}(t)$ suddenly goes back to $0$ at $t=T_d$. Fig.~\ref{figure_4} shows $Q(t)$
and $Q_T(t)$ for the Lorentzian-spectrum phonon subsystem with
different $T_d$ and $V$. While the bias going back to $0$, $Q(t)$
and $Q_T(t)$ show strong oscillations once again, which will
gradually fade away. In the limit of ($t\rightarrow \infty$), the
system restores the equilibrium as initial, $Q(t)$ and $Q_T(t)$
approaches zero and a finite positive value respectively. For
$|V|\geq \omega^0_q$, $Q_T(\infty)$ is a large value that is
approximately proportional to $T_d$. But for $|V|<\omega^0_q/2$,
$Q_T(\infty)$ is almost zero regardless of $T_d$. So if the device
is operated in that range of $|V|$, there will not be much net heat
generation. In particular, as discussed above, for $-\omega^0_q<V<0$,
$Q_T(t)$ can be negative after the first pulse bias, but $Q_T$ is
eventually $\geq 0$ when the system recovers the equilibrium state.
Actually, because of the third law of thermodynamics, it is
prohibited to absorb heat energy from the zero-T phonon subsystem
under the condition of equilibrium, no matter how to control the
bias.

\section{CONCLUSIONS}
\label{section_4}
In summery, the transient heat generation in the QD driven by a
step-like pulse bias is investigated. We find that the suddenly
switched-on bias can induce a heat absorption from the
zero-temperature phonon subsystem. For a strong QD-lead coupling
case, the current-rising process is much quicker than the heat generation,
so the device can finish the electronic operation before the
emergence of heat generation. In addition, for a small bias, the net
heat generation is always very small no matter how to turn on and
off the device.

\section*{ACKNOWLEDGEMENTS}
This work was financially supported by
NSF-China under Grants Nos. 10821403, 10974236, and 11074174,
China-973 program and US-DOE under Grants No. DE-FG02- 04ER46124.

\end{document}